\documentclass[structabstract]{aa}  
\usepackage{graphicx}
\usepackage{txfonts}
\begin{document}
   \title{Period variation in BW Vulpeculae redux}

   \author{Andrew P. Odell}

   \institute{Dept. of Physics and Astronomy, Northern Arizona University
              Flagstaff AZ 86001 USA\\
              \email{Andy.Odell@NAU.EDU}
             }

   \date{Received 15 April 2012; accepted 16 May 2012}

  \abstract
   {}
   {For the past 25 years, BW Vulpeculae has been the topic of period analyses centered on a secular period change with a periodic variation superposed, presumed to be due to light time effects in a binary system.  According to this paradigm, one would expect what seems like a period increase of about 0.5 s during or soon after the year 2001.}
   {I have continued photometric monitoring through the year 2012, adding 35 new timings of maximum and minimum light.}
   {This expected change in period did not occur, which rules out that interpretation of the period variation.  As of 2012, the observed timings are about two hours early compared to those predicted by the quadratic ephemeris, but are very close to those predicted by the linear ephemeris.}
   {In fact, the period has remained constant for the last 32 years, indicating that the previous epochs of constant period are almost certainly the correct interpretation, though the cause of the period changes is still not clear.  Continued photometric monitoring of BW Vul leads to the conclusion that the period changes are abrupt, followed by epochs of constant period lasting between 12 and at least 32 years.}


   \maketitle
%
%________________________________________________________________

\section{Introduction}

Much has been written in the last 30 years about changes in the period of BW Vulpeculae, the monoperiodic $\beta$ Cep star with the largest known amplitude (van der Linden \& Sterken \cite{vander}).  Petrie (\cite{petrie}) first suggested a constant rate of period increase $dP/dt$ = +3.7 s/cent based on spectroscopy, while Cherewick \& Young (\cite{cherewick}) confirmed this with photometry but derived a rate half as large.  These large positive $dP/dt$ would indicate that, if due to the evolution of the star, BW Vul must be in the shell hydrogen burning phase. This contradicts some evidence that $\beta$ Cep stars in clusters are in the late core burning stage, where the rate of period change is well less than 1 s/cent, and it certainly rules out the overall contraction phase where the period should decrease.

Odell (\cite{odell}) noticed what appeared to be a periodic variation superposed on the quadratic ephemeris with half-amplitude about 17 minutes and about 25 year period, and suggested either a light-travel time effect (LiTE) of a small-mass companion or two pulsation modes beating with that long period.  Pigulski (\cite{pigulski}) solved for the binary orbit postulated by Odell, and found a mass function of 0.012 $M_{\sun}$, with a period of 33 years and an eccentricity of 0.46.  For reasonable assumptions about the inclination and mass of the primary, this suggests the secondary is less than 2.5 $M_{\sun}$, so it is undetectable in extant observations.

Tunca (\cite{tunca}) proposed that the period of BW Vul had been constant, but abruptly increased in 1972 by about 0.5 s.  Similar interpretation was made by Chapellier (\cite{chapellier85}) who suggested abrupt period changes in 1931 of +0.61 s, 1945 of +0.50 s, and 1971 of +0.52 s, with the period changes taking place over several years.  He finds that the standard deviation of the residuals for the linear ephemeris fits are reduced by a factor of four over that of the single parabolic ephemeris, but this does not take into account the putative variation due to LiTE, which Pigulski (\cite{pigulski}) shows also substantially reduces the residuals.

Chapellier's (\cite{chapellier85}) suggestion of abrupt period changes followed by epochs of constant period just happens to mimic a constant $dP/dt$ with a periodic variation superposed on it, but the latter makes a very strong prediction that another seeming period increase should have occurred in or around 1980-81, and this exactly happened (Chapellier \& Garrido \cite{chapellier90}).  According to Chapellier's (\cite{chapellier85}) suggestion, this would be fortuitous coincidence, though he proposes no physical mechanism for causing the period to change, but relies on a possibility that some convective process in the star is responsible.  It is unfortunate that the international campaign (Sterken et al. \cite{sterken86}) to monitor BW Vul was organized for the 1982 observing season, and only one timing was done in the year of the period change, 1981.  This led Chapellier \& Garrido (\cite{chapellier90}) to suggest that both the amplitude and timings became unstable for three years during the change.

All this has caused our understanding of BW Vul to be in upheaval, which has been mostly ignored because of the seeming predictive power of LiTE, and so the case has been considered settled (see eg Horvath et al. \cite{horvath}, hereafter HGF).  BW Vul went through more than two complete cycles of LiTE between 1934 and 2001.  Two excellent review papers on the (O$-$C) (residuals) diagram, Sterken (\cite{sterken}) \& Zhou (\cite{zhou}) both use BW Vul as the illustrative example of a star showing LiTE effects.

However, the LiTE model makes another prediction, that there should again be a period change around 2002, with an increase of about 0.5 s.  The purpose of this paper is to report that such a change did \textit{not} happen, and therefore both the large rate of change of period and the LiTE can be ruled out, at least in this star.  The failure of the star to change period again is manifest in a discrepancy by 2012 of over two hours compared to the prediction of the cannonical model, as shown in Section 2.  Though this does not confirm the piecewise-linear ephemeris with abrupt period changes, it is consistent with it; continued monitoring might clarify the situation in this regard.

%__________________________________________________________________

\section{Observations}

BW Vul was monitored at Lowell Observatory from 1995 to 2011 using three different instruments.  Before 2002, the white photoelectric photometer was used on the 31-inch telescope.  On JD 52948 and 55844, the NASACam CCD was used in robotic mode on the 31-inch telescope (\textit{note} all HJDs have 2,400,000 subtracted).  The rest of the measurements reported here were obtained with the photoelectric photometer permanently mounted on the 21-inch telescope.  Primarily the Str\"omgren b and y filters were used, though others were employed occasionally.  In all cases the comparison star was HD198820=HR7996, which is similar color and brightness to the variable, so no color terms were included.  The comparison was used to find the extinction coefficient, and all measurements were corrected for extinction; the comparison was interpolated to the time of the variable measure, and the magnitudes were subtracted.

These observations are summarized in Table~\ref{table:1}.  Col. 1 gives the cycle number based on the maximum of Huffer (\cite{huffer}) being cycle 0.  Rather than convert times of minimum, cycles with fraction 0.443 were assigned (Sterken et al \cite{sterken87}).  Col. 2 gives the HJD of the timing - this was found by folding back in time the light curve so the rising and falling branches coincided; the time at the fold was taken as time of maximum or minimum, thus making obvious any effects of the stillstand.  The fourth column is the residual, or observed time minus the computed time (O$-$C) based on a linear ephemeris fit to timings between 1982 and 1998 (Eq. 4); these are plotted in Fig. 4.  The fifth column is (O$-$C) based on an ephemeris which includes a quadratic fit (Eq. 2) between 1934 and 2001 (two complete cycles of the supposed LiTE) and a removal of the LiTE (Eq. 3); these are plotted in Fig. 3.

\textit{Note} that neither ephemeris utilizes the potentially controversial timings after 2002.  The timings used for the fits and the graphs include supplemental times from Sterken (priv. comm.), HGF, and the AAVSO.  The entire list of 335 timings is available from the author.

%                                             Table 1
%_____________________________________________________________
%
\begin{table} 
\caption{Timings for BW Vulpeculae}                % title of Table
\label{table:1}                                    % is used to refer this table in the text
\centering                                         % used for centering table
\begin{tabular}{l c c r r}                       % centered columns (6 columns)
\hline\hline                                       % inserts double horizontal lines
     Cycle &     HJD (2400000+)  & Max/Min & (O$-$C)\textsuperscript{1} & (O$-$C)\textsuperscript{2} \\   % table heading 
\hline                                                  % inserts single horizontal line
    105517.443 & 50015.6330 &  Min &   2.2 &    3.1 \\  %   1995 Oct 25 - odell, rodriquez 
    105523.    & 50016.7540 &  Max &   8.8 &    9.7 \\  %   1995 Oct 26 - odell
    107209.    & 50355.7087 &  Max &   1.1 &    2.9 \\  %   1996 Oct 29 - odell
    108568.    & 50628.9300 &  Max &   4.9 &    6.8 \\  %   1997 Jun 29 - odell, shutts good
    108572.    & 50629.7340 &  Max &   4.6 &    6.5 \\  %   1997 Jun 30 - odell         good
    109134.    & 50742.7213 &  Max &   5.5 &    7.3 \\  %   1997 Oct 21 - odell, shutts  ave of B&V good night
    111069.    & 51131.7371 &  Max &  $-$0.5 &   $-$0.2 \\  %   1998 Nov 13 - odell - good night, one color
    116460.443 & 52215.6610 &  Min &   8.5 &   $-$3.7 \\  %   2001 Nov 02 - ODELL - min
    116461.    & 52215.7757 &  Max &  13.6 &    1.4 \\  %   2001 Nov 02 - odell  airmass > 2
    116465.    & 52216.5737 &  Max &   4.7 &   $-$7.5 \\  %   2001 Nov 03 - odell
    116480.    & 52219.5990 &  Max &  18.6 &    6.3 \\  %   2001 Nov 06 - odell/pick
    116485.    & 52220.6055 &  Max &  20.4 &    8.1 \\  %   2001 Nov 07 - odell/sackey
    117362.    & 52396.9143 &  Max &  10.7 &   $-$4.7 \\  %   2002 May 02 - odell/tse
    117615.    & 52447.7815 &  Max &  15.2 &   $-$1.3 \\  %   2002 Jun 22 - odell/sackey
    117616.    & 52447.9690 &  Max &  $-$4.3 &  $-$20.8 \\  %   2002 Jun 22 - odell/sackey   off 13 min from linear; 7 pts
    117620.    & 52448.7823 &  Max &   8.8 &   $-$7.6 \\  %   2002 Jun 23 - odell/prescott  
    117695.    & 52463.8650 &  Max &  15.2 &   $-$1.6 \\  %   2002 Jul 08 -    mediocre
    120106.    & 52948.5758 &  Max &   6.4 &  $-$20.9 \\  %   2003 Nov 05 - Buie CCD, estimate   y .5758, b .5758
    120106.443 & 52948.6670 &  Min &   8.3 &  $-$19.0 \\  %   2003 Nov 05 - Buie CCD  min        y .6665  b .6675
    120107.    & 52948.7879 &  Max &  22.3 &   $-$5.0 \\  %   2003 Nov 05 - Buie CCD  max, high airmass  y .7880  b .7878
    127423.    & 54419.6259 &  Max &  23.3 &  $-$47.7 \\  %   2007 Nov 15 - 21" - pretty good
    134084.    & 55758.7718 &  Max &  12.3 & $-$112.3 \\  %   2011 Jul 16 - 21"
    134084.443 & 55758.8585 &  Min &   7.7 & $-$116.8 \\  %   2011 Jul 16 - minimum - 21"
    134188.    & 55779.6750 &  Max &   4.6 & $-$120.9 \\  %   2011 Aug 06 - 21"
    134188.443 & 55779.7662 &  Min &   6.5 & $-$119.0 \\  %   2011 Aug 06 - 21' minimum
    134193.    & 55780.6825 &  Max &   7.8 & $-$117.7 \\  %   2011 Aug 07 - 21"  good night
    134193.443 & 55780.7708 &  Min &   5.6 & $-$120.0 \\  %   2011 Aug 07 - 21" minimum
    134263.    & 55794.7563 &  Max &   8.9 & $-$117.3 \\  %   2011 Aug 20 - 21" some clouds,  y 0.7560, b 0.7565
%    134427.    & 55827.7157 &  Max &  -8.1 & -135.8 \\  %   2011 Sep 23 - 31" Robo CCD, only before max
    134446.443 & 55831.6397 &  Min &  12.5 & $-$115.3 \\  %   2011 Sep 27 - 21" not a great night
    134447.    & 55831.7505 &  Max &  11.9 & $-$115.9 \\  %   2011 Sep 27 - 21" again, not great night
    134511.    & 55844.6161 &  Max &  10.2 & $-$118.2 \\  %   2011 Oct 10 - Robo - stdev 0.0005d = 0.74m
    134511.443 & 55844.7042 &  Min &   7.7 & $-$120.8 \\  %   2011 Oct 10 - Robo - stdev 0.0006d = 0.93m
    134512.    & 55844.8123 &  Max &   3.2 & $-$125.2 \\  %   2011 Oct 10 - Robo - high am, looks OK, stdev 0.0003d = 0.43m
    135542.    & 56051.8905 &  Max &   7.5 & $-$130.4 \\  %   2012 May 4  - Robo - high am  sd 0.0014
    135542.447 & 56051.9803 &  Min &   7.4 & $-$130.5 \\  %   2012 May 4  - Robo - min      sd 0.0010

\hline                                             %inserts single line
\end{tabular}
(O$-$C)\textsuperscript{1} is the residual to a linear fit of data between 1981 and 2002.
(O$-$C)\textsuperscript{2} is the residual to a quadratic fit plus LiTE model (HGF Eq.8).
\end{table}

\section{Analysis}

The method of analysis adopted here is to utilize the exact steps of HGF.  Theirs is the most recent attempt to understand the period behavior of BW Vul, including data through 1997, and it reproduces the results of several earlier investigations.  That paper also gives all the fitting functions which result from the secular period increase and LiTE.

%
%                                                Two column figure
%-----------------------------------------------------------
   \begin{figure}[ht]
   \centering
   \includegraphics[width=9cm]{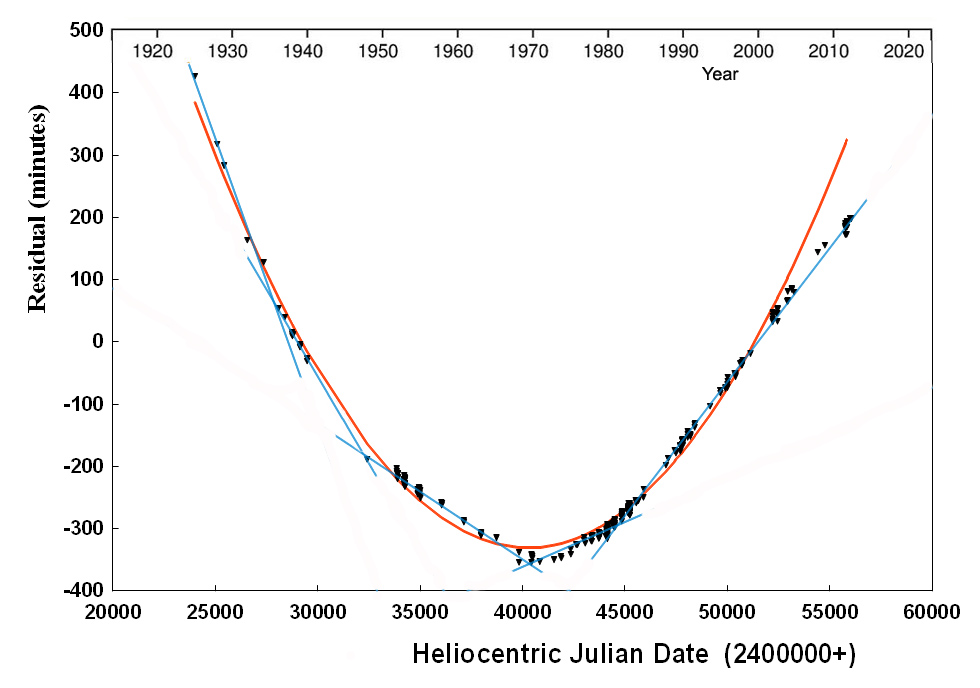}
      \caption{Residuals for HGF Eq 2 (my Eq 1), the solid lines are fits of a parabola and five piecewise-linear fits to subsets of the data.}
         \label{f1}
   \end{figure}

The first step in analyzing the period of a variable star is to fit a linear ephemeris to the timings \textit{vs} cycle number; the slope of this line would be the period.  Then, the (O$-$C) diagram (residuals plotted \textit{vs} HJD) will show how well that ephemeris works, and whether there is some trend away from it.  When this is done for BW Vul, the resulting linear equation (HGF Eq. 2, taken from Sterken (\cite{sterken93}))  \footnote{I have changed the initial epoch to reflect that I use timings of maximum light for all their equations.  This is because most historical timings are time of maximum light. I make a plea here to modify the cycle count rather than the actual timings, as I've done in Table~\ref{table:1}.  This, I believe, is more academically honest, as it doesn't require modifying the data.  Likewise, I use actual times of radial velocity zero crossing, with a phase offset by 0.522 in the cycle count rather than infer a time of maximum or minimum light from RV curves.} is:
\begin{equation}
   HJD\mathrm{_{max}} = 28802.5487 + 0.201038~E
\end{equation}
which gives the residuals shown in Fig.~\ref{f1}.  It is clear from that figure that a single constant period doesn't represent the timings over the entire dataset (there are residuals up to $\pm 400 minutes$), but the residuals exhibit what appears to be a parabola.  The residuals, however, seem to show a periodic excursion on either side of the parabloic fit, with amplitude about 20 minutes and period about 35 years.  Note that the residuals can also be fit with a series of five straight lines.  These are the two paradigms mentioned in the introduction.  Note too, that after HJD 53000, the residuals fall further and further below the parabola which fit so well for 70 years, but the observations continue along the fifth of the straight line segments.

\begin{figure}[ht]
   \centering
   \includegraphics[width=9cm]{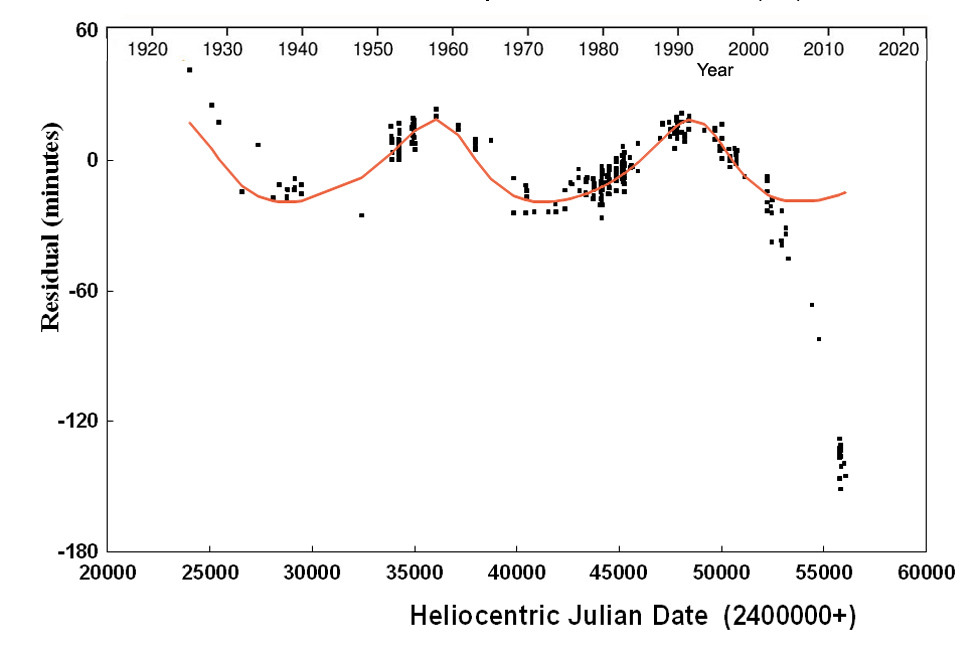}
      \caption{Residuals for HGF Eq 6 (my Eq 2), a quadratic fit to all data.  The solid line is the residuals expected from LiTE, HGF Eq 8 (my Eq 3)}
         \label{f2}
   \end{figure}

\begin{figure}[ht]
   \centering
   \includegraphics[width=9cm]{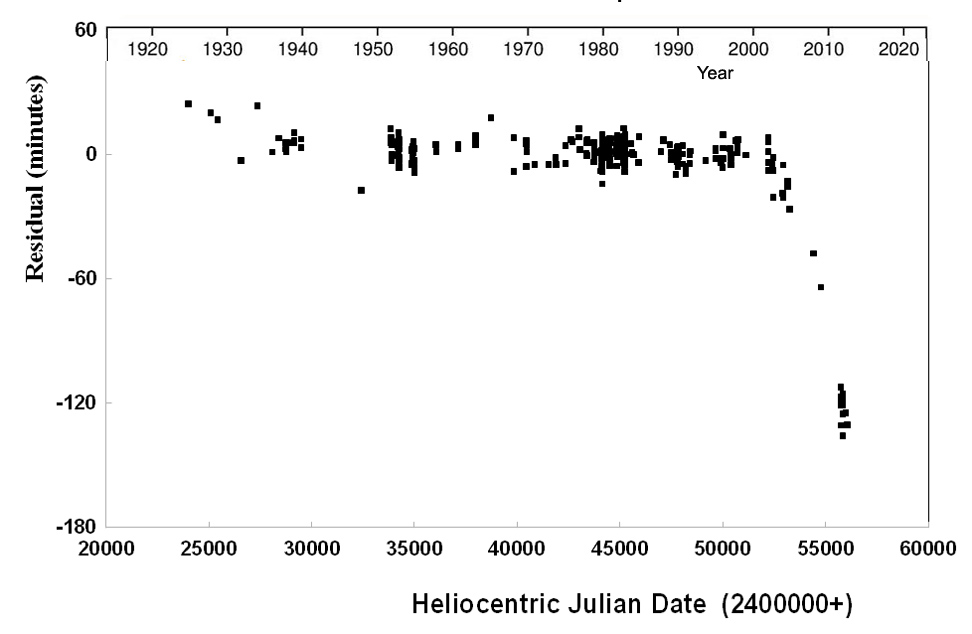}
      \caption{Residuals after removing HGF Eq 8 (my Eq 3), the LiTE fit.}
         \label{f3}
   \end{figure}

Fitting the timings with a parabola requires some care because of the periodic excursion on either side of that curve.  It is thus important to use only a whole number of cycles to minimize the effect of the excursions on the parabola.  The years 1934 to 2001 meet this requirement and have adequate timings; when a quadratic least squares fit is made to just those years, the resulting ephemeris is
\begin{equation}
   HJD\mathrm{_{max}} = 28802.5661 + 0.20102934~E + 0.755\times10^{-10}~E^{2}
\end{equation}
which is HGF Eq. 6 (note that their equation has an error in the quadratic term exponent). Fig.~\ref{f2} shows the residuals after removal of the parabolic fit from Eq. 2.  It also shows the fit to those residuals presented in HGF Eq. 8:
\begin{equation}
  (O$-$C)\mathrm{_{LiTE}} = \textit{A}~\left[~{0.7791~\sin(\Phi+5.274) \over 1.0+0.47~\cos\Phi} - 0.3986~\right]
\end{equation}
where \textit{A} = 0.013630223 in days and $\Phi = 0.00010084(E-7934)$ in radians. This function is plotted in Fig.~\ref{f2} and when subtracted from those residuals, the new ones are plotted in Fig.~\ref{f3}.  In both of those figures, the increasingly poor agreement after HJD 53000 can easily be seen, and amounts to more than two hours by 2012.

%
%______________________________________________________________

%                                                Two column figure
%-----------------------------------------------------------
%
%______________________________________________________________

%                                                Two column figure
%-----------------------------------------------------------

The other paradigm for the period variations in BW Vul, proposed by Tunca (\cite{tunca}) and amplified by Chapellier \& Garrido (\cite{chapellier90}), is that the period remains constant for one or more decade, and then undergoes an abrupt period change.  To test the new timings against this model, a linear ephemeris was fit to the data from 1982 to 1998. which yielded an ephemeris:
\begin{equation}
   HJD\mathrm{_{max}} = 28801.9903 + 0.2010439~E 
\end{equation}
Recall that the cycle count for this equation is still based on a cycle count where the Huffer {\cite{huffer} maximum is considered to be cycle zero.  It is important, when applying an ephemeris function to new data to not include the new data in the least squares fit.  If it is included, a period change will be masked for several observing seasons because the least squares fit will automatically adjust to the new, slightly disparate timings.

Fig.~\ref{f4} shows the residuals of this ephemeris to the fitted points (diamonds) and the new points (squares) after HJD 52000 (year 2000).  Though the new points seem to have slightly positive residuals, there is no net trend away from zero, so no new period change is indicated.  While the continuation of the linear ephemeris does not guarantee this second interpretation to be correct, it is certainly consistent with it.  This brings up the question of how the transition in period takes place.

%______________________________________________________________

%                                     Two column figure (place early!)
%______________________________________________ 
\begin{figure}[ht]
\centering
\includegraphics[width=9cm]{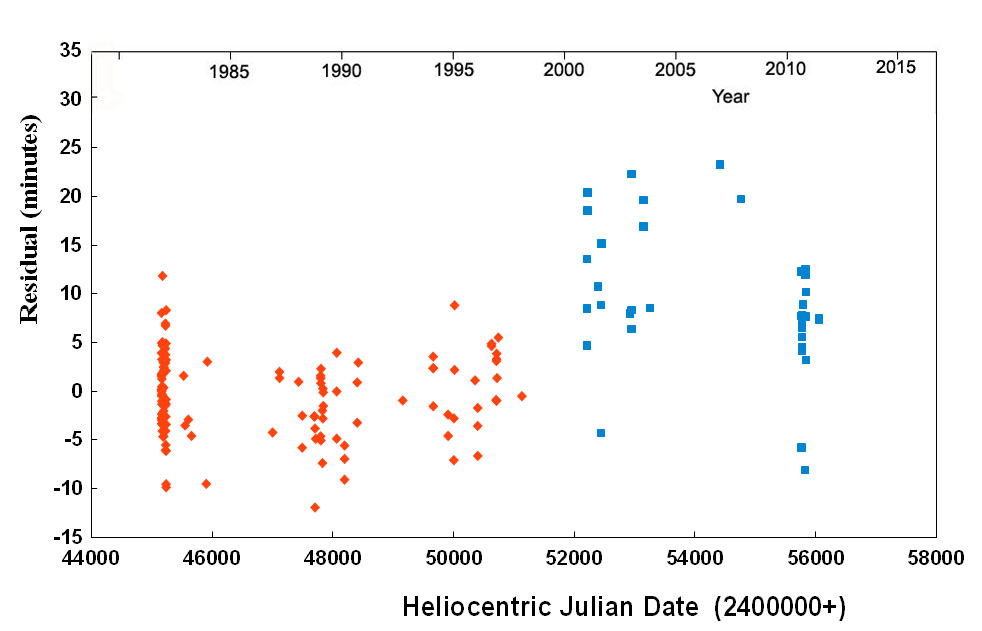}
\caption{Residuals from the linear fit given in Eq 4.}
           \label{f4}%
\end{figure}

Chapellier \& Garrido (\cite{chapellier90}) analyze the 1981 period change and conclude that it showed a transient phase over the years 1979 to 1982 (about three years), with cycle-to-cycle variations of both amplitude and phase.  However, they present only four sparse light curves, one consisting of only six measurements.  Their Fig. 2 shows a large scatter in the residuals from 1981, but this was based on just one measurement.  

Fortuitous good luck brought to light additional data obtained by James Kemp in 1981 when he was doing polarimetry of BW Vul, but neither the polarimetry nor the concommitant photometry were ever published.  The photometry is not optimal, as a rather faint nearby star was used as a comparison, but it does allow eight new timings (listed in Table~\ref{table:2}) to be derived for a critical year in the transition that previously had only one timing.  These observations were communicated by Dr. Gary Henson.

\begin{table} 
\caption{Timings for BW Vulpeculae from James Kemp 1981}    % title of Table
\label{table:2}                                             % is used to refer this table in the text
\centering                                                  % used for centering table
\begin{tabular}{l c c r }                                   % centered columns (6 columns)
\hline\hline                                                % inserts double horizontal lines
     Cycle &  HJD (2400000+)   & Max/Min & (O$-$C)\textsuperscript{1}  \\   % table heading 
\hline                                                  % inserts single horizontal line
 79793.     &   44843.8917  & Max &  9.9 \\     % x   1981 Aug 27 - Kemp B HJD
 79798.     &   44844.8982  & Max & 11.8 \\     % x   1981 Aug 28 - Kemp B HJD
 79803.     &   44845.8962  & Max &  1.4 \\     % 
 79807.     &   44846.7097  & Max & 14.8 \\     % x   1981 Aug 30 - Kemp Red HJD
 79813.     &   44847.9131  & Max & 10.7 \\     % x   1981 Aug 31 - Kemp B HJD
 79818.     &   44848.9161  & Max &  7.5 \\     % x   1981 Sep 01 - Kemp B HJD
 79887.     &   44862.7868  & Max &  5.6 \\     % x   1981 Sep 15 - KEMP B HJD
 79887.443  &   44862.8838  & Min & 17.0 \\     % m   1981 Sep 15 - Kemp B HJD minimum
\hline                                             %inserts single line
\end{tabular}
\\(O$-$C)\textsuperscript{1} is the residual to the linear ephemeris of Eq. 4.
\end{table}

Fig.~\ref{f5} shows an enlarged and extended view of Fig.~\ref{f4} between 1976 and 1984.  The diamonds are part of the timings that were fit by Eq. 4; the squares were not fit at all, but lie along a straight line which represents the shorter period prior to 1980; and the triangles represent timings in the 1980, 1981, and 1982 seasons (also not fit).  While it seems as though they lie on a curve above the straight line fits, they are only above by a few minutes, and it isn't clear if this is significant; certainly the scatter is almost as large as the offset itself.  Further, the scatter in those years is comparable to the scatter in prior and subsequent years.  There doesn't seem to be the instabilities in period that Chapellier \& Garrido (\cite{chapellier90}) have claimed.  Their suggestion that the amplitude was also varying by 15\% could be due to inhomogeneous data from different filters as well.  There was no variation at that level seen in the light curves of the international campaign of Sterken et al (\cite{sterken87}), which was also in the transition years.  

%                                     Two column figure (place early!)
%______________________________________________ 
   \begin{figure}[ht]
   \centering
   \includegraphics[width=9cm]{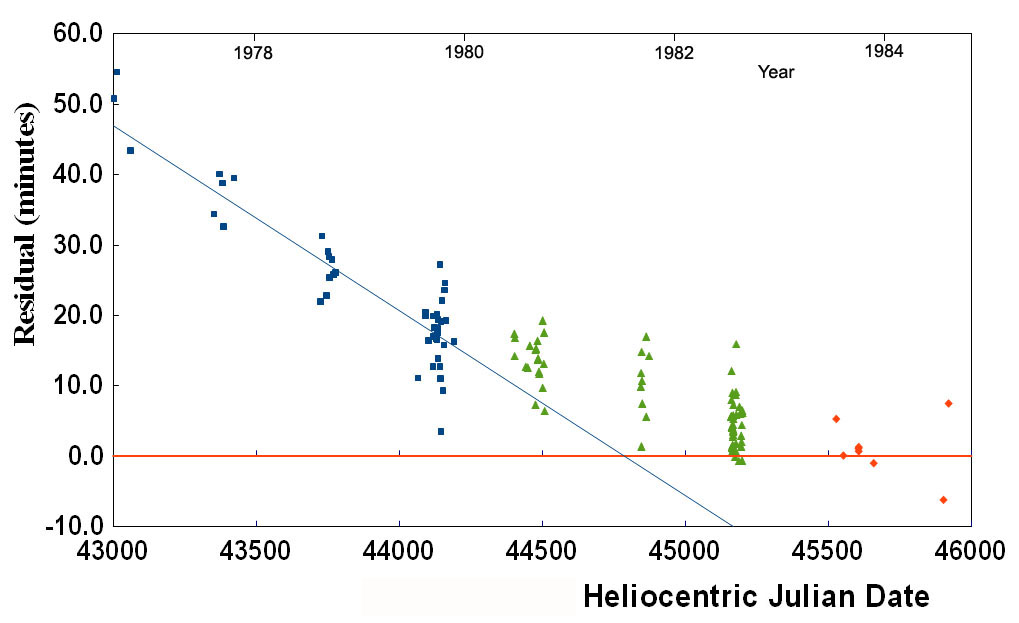}
   \caption{Residuals from the time of the 1981 period change, based on the ephemeris in Eq. 4}
              \label{f5}   %
    \end{figure}

\section{Conclusions}

Photometry of BW Vulpeculae since 1995 fails to verify the continued secular period change or the cyclic period variation attributed to the LiTE of a binary system.  It appears that the proper interpretation of the period variation in this star is a piecewise linear ephemeris, \textit{ie} constant period interrupted each one to three decades by an abrupt period change.  If the discrete period changes are not evolutionary, this allows the star to be in the core hydrogen burning phase of evolution, consistent with the position of cluster members of this class of star in the color-magnitude diagram.  It thus seems to help solve the problem of an inordinate fraction of $\beta$ Cep stars in the shell-burning phase, as pointed out by Jerzykiewicz (\cite{jerzy}).

It is anticipated that another period change should take place in the not-too-distant future, so continued monitoring of the timing and light curve could prove to be very important.

\begin{acknowledgements}
I would like to thank Lowell Observatory for providing access to their telescopes, instruments and library, and especially Wes Lockwood who helped with that access.  I appreciate the advice, encouragement, and unpublished timings from Christiaan Sterken, and James Kemp's unpublished data which was made available by Gary Henson.  I thank Eric Chapellier for helpful comments on the manuscript.  This work made use of the SIMBAD database, operated at CDS, Strasbourg, France
\end{acknowledgements}

\end{document}